\documentclass[journal]{IEEEtran}

\ifCLASSINFOpdf
  \usepackage[pdftex]{graphicx}
\else
\fi

\usepackage{amsmath}
\usepackage{import}
\usepackage{bm}
\usepackage{cite}
\usepackage{amsfonts}
\usepackage{amssymb}
\usepackage{color}

\ifCLASSOPTIONcompsoc
\else
  \usepackage[caption=false,font=footnotesize]{subfig}
\fi

\usepackage{booktabs}
\newcommand{\ra}[1]{\renewcommand{\arraystretch}{#1}}

\begin{document}

	\newcommand{\keyw}[1]{{\bf #1}}
	\newcommand{\reffig}[1]{figure.\,\ref{#1}} 
	\newcommand{\refeq}[1]{Eq.\,(\ref{#1})} 
	\newcommand{\refsec}[1]{Section \ref{#1}} 
	\newcommand{\N}{\mathbb{N}} 
	\newcommand{\R}{\mathbb{R}} 

\title{To Recurse or not to Recurse,\\
a Low Dose CT Study}

\author{Shabab Bazrafkan,
	Vincent Van Nieuwenhove,
	and Jan Sijbers 
	\thanks{S. Bazrafkan, and J. Sijbers are with imec Visionlab, Department of Physics, University of Antwerp,
		Antwerp, Belgium e-mail: \{shabab.bazrafkan\},\{jan.sijbers\}@uantwerpen.be.}
  \thanks{V. Van Nieuwenhove is with Agfa NV, Mortsel, Belgium
		email: vincent.vannieuwenhove@agfa.com.}}


\maketitle

\begin{abstract}
	Restoring high-quality CT images from low dose CT counterparts is an ill-posed, nonlinear problem to which Deep Learning approaches have been giving superior solutions compared to classical model-based approaches. In this article, a framework is presented wherein a Recurrent Neural Network (RNN) is utilized to remove the streaking artefacts from low projection number CT imaging. The results indicate similar image restoration performance for the RNN compared to the feedforward network in low noise cases while in high noise levels the RNN return better results. The computational costs are also compared between RNN and feedforward networks. 
\end{abstract}


%
\IEEEpeerreviewmaketitle

\section{Introduction}

Recurrent Neural Networks (RNN) are a set of  deep learning approaches wherein the temporal information of the input signal is taken into account while processing the current timeslot. In other words, RNNs contain a memory also known as the \textit{state} which changes based on their previous state, previous output, and the current input signal. An RNN structure allows to store, forget, or pass  information from remote time slots into the current one \cite{RNN1}. These networks are used in a large number of applications including Natural Language Processing (NLP) \cite{NLP1,NLP2,NLP3}, video processing \cite{RNNVIDEO1,RNNVIDEO2}, trajectory prediction \cite{TRAJECTORY1}, and correlation analysis \cite{CORRELATION1}.\\

The early versions of RNNs suffer from the gradient vanishing due to the {\it sigmoid} and {\it tanh} nonlinearities used in the network which also made long term dependencies nontrivial to learn. With the introduction of Long Short Term Memories (LSTM) \cite{LSTM}, this issue was resolved. These processing units take advantage of gated architecture wherein the passing or ignoring of the information flow is decided by several blocks such as `forget', `input' and `output' gates. These units provide the opportunity to learn the long-distance correlations between remote time slots. This is an important quality for speech recognition, text processing, and NLP.\\
 
In 2014, a new recurrent processing block known as Gated Recurrent Unit (GRU) was introduced \cite{GRU} wherein, despite its simpler architecture compared to LSTM, it delivers comparable outcomes. These processing units are widely used in recurrent network design and the convolutional counterpart of these units are investigated in \cite{CONVRNN}. 
The RNNs are also used in medical applications. In \cite{RNNMED1}, RNNs are used in sequence labeling in the unstructured text of Electronic Health Record (EHR) notes which have critical health-related applications such as pharmacovigilance and drug surveillance. In \cite{RNNMED2}, a similar approach is utilized for prognostication and determining life expectancy from the EHR notes.\\
 
In \cite{RNNMED3}, the authors present a user physical activity predictor based on wearable sensors using the RNN approach. In \cite{RNNMED4},  a method is presented to predict lung cancer treatment over time by investigating the cancer progression, distant metastases, and local-regional recurrence using a mixture of CNN and RNN taking advantage of GRU blocks.
The work presented in \cite{RNNMED5}, uses an RNN based Generative Adversarial Network (GAN) to train a deep generator for Electrocardiogram signal. The proposed network takes advantage of both CNN and LSTM recurrent units.\\ 

The recurrent scheme is also applied to convolutional networks to provide higher quality outputs while preserving spatial information. In \cite{RNNMED6}, the convolutional recurrent units are used to acquire high-grade segmentation maps for cardiac MRI images using an architecture called recurrent interleaved attention network, which processes the input image in different pooling scales in a recurrent scheme.\\
  
In \cite{RNNMED7}, an RNN is used in conjunction with the level set idea wherein the network returns an evolved level set function in each iteration. This method is used for biomedical image segmentation use cases. Another approach presented in \cite{RNNMED8} uses the well-known U-net \cite{UNET} architecture in a recurrent scheme to provide segmentation for medical images. In the current article, convolutional GRUs are used as the main recurrent processing units to remove the streaking artefact from the low dose CT images; explained below.\\

Computed Tomography (CT) is a well-known, non-invasive imaging technique to visualize the interior of object. The basic concept of CT relies on an X-ray source and a detector unit synchronously rotating around the object, while taking radiographs from each angle. The acquired set of radiographs is then preprocessed after which  a 3D image is reconstructed using mathematical reconstruction models such as filtered backprojection or algebraic reconstruction techniques.\\

The quality of the reconstructed 3D CT image strongly depends on the number of acquired radiographs during scanning. The denser the angular sampling during rotation of the gantry and the higher the intensity of the X-ray source, the better the CT image quality is. However, the number of radiographs and the X-ray intensity during the acquisition of a radiograph is directly related to X-ray dose, which ought to be as low as possible, especially in biomedical applications. A straightforward way to lower the dose in X-ray CT is to lower the number of acquired radiographs during the scan. However, this leads to undersampling artefacts in the reconstructed image. Many methods have been proposed in the literature to reduce those artefacts, mainly based on computationally expensive iterative reconstruction methods. With the uprising of DNNs, promising deep learning approaches have been introduced for low dose CT.\\

In the current study, a Recurrent Neural Network is utilized to remove the streaking artefact of limited projection CT reconstructions. The main motive for using RNN comes from the fact that the streaking artefacts look similar in different parts of the reconstructed image which allows learning such artefacts from a patch of the image and apply it to some other distant patches. To the best of our knowledge no recurrent scheme has been exploited in removing low dose artefacts from the CT reconstruction. The results are also compared to the feedforward network as well as model-based approaches.\\

In the next section, the physics of the CT imaging is discussed followed by explaining the common reconstruction methods. The recurrent neural networks are also explained in the next section. Network design is covered in section \ref{sec:NetDesign} and section \ref{sec:Database} is dedicated to the databases used in this study. The training procedure is explained in section \ref{sec:training} and results and conclusions are presented in sections \ref{sec:results} and \ref{sec:conslusions} respectively.
\section{Materials}
\label{sec:materials}
\subsection{CT Imaging}

Computed tomography (CT)  is well-known imaging technique that allows for non-invasive visualization of the interior of an object.  It is widely used in many applications such as medical imaging \cite{Bach2012,Kubo08}, non-destructive testing \cite{DeSchryver2016}, industrial metrology \cite{HILLER2016}, food industry \cite{Schoeman2016,Janssens2015}, and security \cite{SHIKHALIEV2018}.
In CT,  X-ray radiation is used to acquire a number
of two dimensional (2D) images of an object from many different view points.
From these images, cross-sections (tomograms) of the object's internal structure are computed using a reconstruction algorithm and subsequently analyzed.
In this section, we will shortly describe the principle of X-ray CT imaging and image reconstruction. 

\subsection{X-rays: matter interaction and detection}
\label{sec:introduction:acquisitionProcess:Xray}



When an X-ray beam passes through an object, its intensity decreases due to physical mechanisms such as the photo-electric effect or elastic or inelastic scattering. 
Let $I_0$ denote the intensity of a monochromatic X-ray beam that leaves the X-ray source. 
Then the intensity of the X-ray beam at position $s$ on the detector after passing through the object along a line $L$ oriented at angle $\theta$  is given by:
\begin{equation}
I_\theta(s) = I(0) e^{-\int_L \mu(\eta)d\eta} . \label{eq:1:derBeerLambert10}
\end{equation}
\refeq{eq:1:derBeerLambert10} describes the relationship between the observed intensity at the detector side and the unknown attenuation coefficients $\mu$ the X-ray beam passed through. Log-normalization of this detected intensity yields the projection value
\begin{equation}
p_\theta(s) = -\ln\left(\frac{I_\theta(s)}{I_0}\right) =
\int_L \mu(\eta)d\eta .
\label{eq:1:BeerLambertLinearForm}
\end{equation} 
which linearly relates to the (unknown) attenuation coefficients of the object. 

The main purpose of CT reconstruction methods is to recover the object's attenuation coefficients $\mu(.)$ from given projection data $p(.)$. In what follows, we describe commonly used reconstruction methods to recover the object's attenuation values from  projection data $p_\theta(s)$ measured by directing the X-ray beam at different angles $\theta$ through the object. 


\subsection{Analytical reconstruction methods} \label{sec:introduction:reconstructionMethods:analytical}
In the analytical approach, the object's attenuation distribution is described as a function $f: \mathbb{R} \times \mathbb{R} \rightarrow \mathbb{R}$ that maps the spatial coordinate $(x,y)$ to its corresponding local attenuation coefficient $\mu$. 

The Filtered Back Projection (FBP) is a commonly used analytical reconstruction method, based on the following analytical formula:
\begin{equation}
f(x,y) = \int_0^{\pi}\left\{ \int_{-\infty}^{\infty}P_{\theta}(q)|q|e^{2\pi i q (x \cos\theta + y\sin\theta)}dq \right\} d\theta .
\label{eq:1:FBPformula}
\end{equation}
where $P_{\theta}(q)$ denotes the Fourier transform of the projection data $p_{\theta}(s)$. 
As can be observed from \refeq{eq:1:FBPformula}, the FBP formula gives rise to a two step approach for calculating a reconstruction of the scanned object based on the measured projection data \cite{VanEyndhoven2018}: 1) Filter the Fourier transform of the projection $p_{\theta}(s)$ by a ramp filter to account for the radial sampling in Fourier space. 2) For a particular pixel $(x,y)$, sum up all the filtered projection data that corresponds to the lines $x \cos\theta + y\sin\theta$ with $\theta \in [0,\pi]$. 

The analytical representation (\ref{eq:1:FBPformula}) assumes that projection data is available from a sufficiently large and densely sampled angular range. Under this condition, FBP generally leads to high quality reconstructions.  In case of limited angle scanning or related missing data problems, severe streaking artefacts appear in the reconstructed image.

\subsection{Algebraic reconstruction methods} 
\label{sec:introduction:reconstructionMethods:algebraic}
\begin{figure}[h!]
	\centering
	\def \svgwidth{0.43\textwidth}
	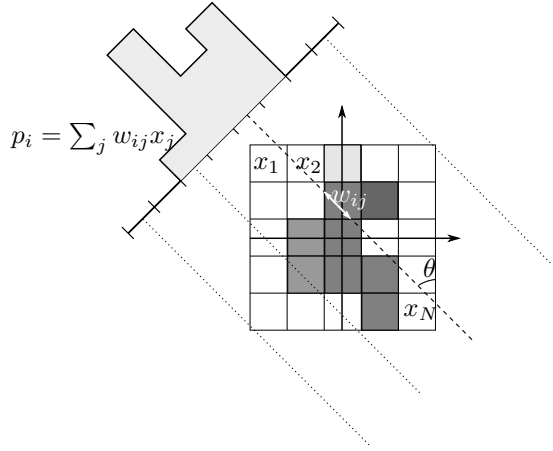
	\caption{The Discrete Projection Model for a parallel beam.}
	\label{fig:1:ProjModelDiscrete}
\end{figure}

Another popular class of reconstruction methods are algebraic reconstruction methods. These  methods rely on a discrete model of the object $\bm x=\{x_j\}$, as shown in \reffig{fig:1:ProjModelDiscrete}. Their basic scheme is to iteratively minimize the difference between the computed forward projection of the discrete image $\bm{p}=\{p_i\}$ with the observed projection data $\bm {\tilde p}=\{\tilde p_i\}$, where the object is updated based on the backprojected difference. Thereby,  $p_i = \sum_j w_{ij}x_j$, with $\bm W={w_{ij}}$ denoting the contributions of object pixel $x_j$ to the detector pixel $p_i$.

A commonly used algebraic reconstruction method is the SIRT algorithm, in which, in each iteration $k$, the current estimate of the image, $\{x_j^k\}$, is updated as follows: 
\begin{equation}
x_j^{(k+1)} = x_j^{(k)} + \frac{1}{\sum_{i} w_{ij}}\sum_{i}\left( \frac{w_{ij}(p_i-\sum_{h}w_{ih}x_h^{(k)})}{\sum_{h} w_{ih}}\right) 
\label{eq:1:SIRTpixelwise}
\end{equation}

\begin{figure}[!ht]
	\centering
	\includegraphics[width=0.45\textwidth]{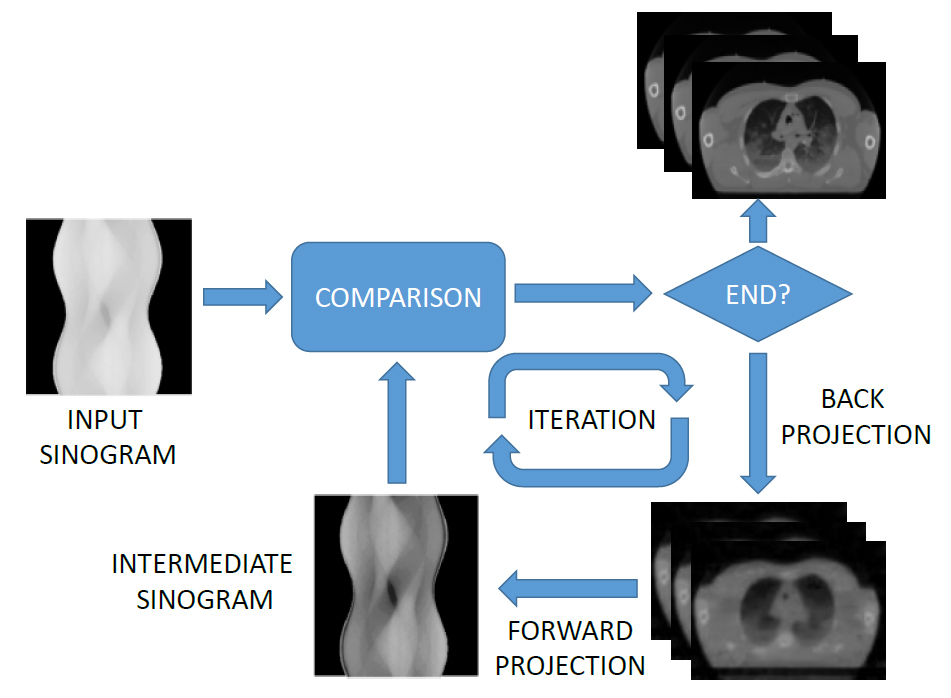}%
	\caption{Iterative reconstruction scheme.}
	\label{fig:iter}
\end{figure}

An important advantage of algebraic reconstruction methods is that, if prior knowledge is available, this information can be easily integrated in such an iterative reconstruction scheme (cfr. Fig. \ref{fig:iter}).

As a result, the quality of the reconstructed image can substantially be improved compared to FBP. Another advantage is that data consistency, being the minimization of the difference between the forward projection of the object $\{x_j\}$ with the measured projection data $\{\tilde p_i\}$, is explicitly enforced. 
One of the most important drawbacks, however, is their computational load and slow convergence, which is the main reason why they are not commonly used in industrial applications.
\\

\subsection{Feedforward vs Recurrent Neural Networks}
\begin{figure}[!t]
\centering
\includegraphics[width=1.6in]{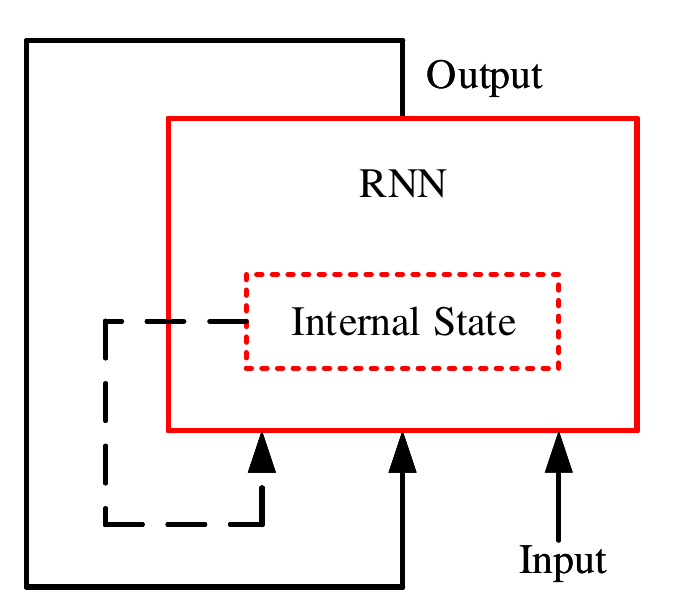}
\caption{Recurrent unit accepts an input, previous timestep output, and internal state.}
\label{fig:RNN1}
\end{figure}

Feedforward neural networks are vastly used in machine learning solutions where the output of the network solely depends on the current input. In other words, these networks do not consist of any internal memory and they do not remember their previous states. They are used in still signal processing including both classification and regression problems. But these architectures are not suitable for addressing every scenario like the problems where a signal changes by time and/or the output at the current stage is highly correlated with the input and output of the previous time slots. For example, in  Natural Language Processing (NLP) use cases, the decisions at each step are related to the words in previous time steps. These problems are handled with a set of neural networks known as Recurrent Neural Networks (RNN) (see figure \ref{fig:RNN1}). In general, recurrent units in each step accept the input of that time step, the output, and the internal state of the previous step and return the output and internal state for the next timestep.\\ 

\begin{figure*}[!t]
\centering
\includegraphics[width=7in]{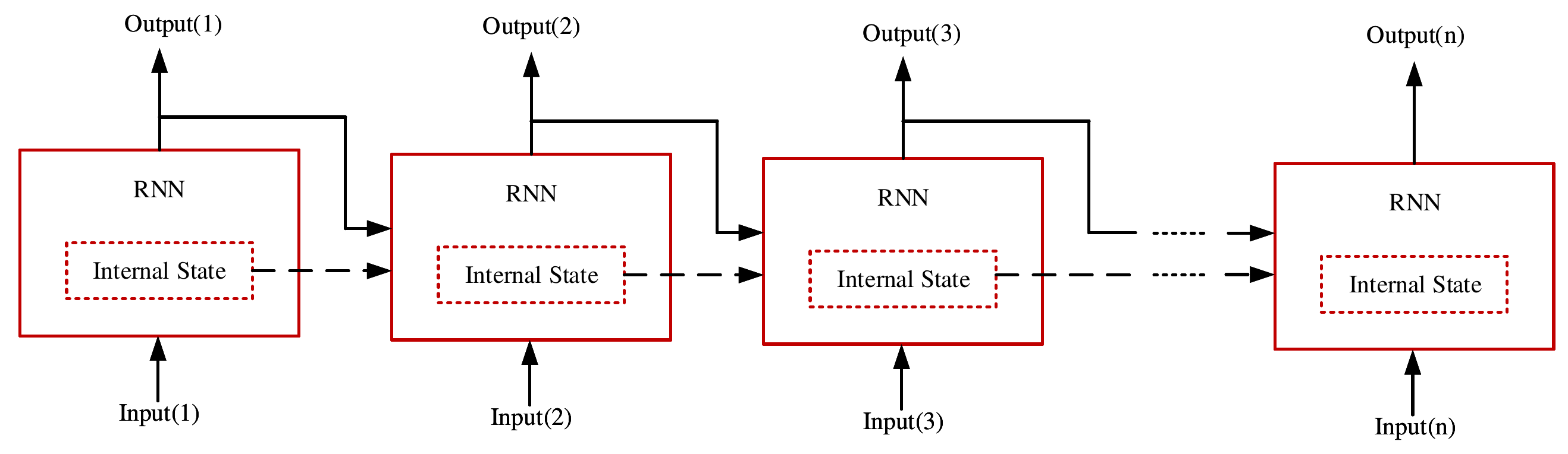}
\caption{The unrolled schematic of the network in Figure \ref{fig:RNN1}.}
\label{fig:RNNunrolled}
\end{figure*}

These networks could be considered as feedforward networks with an added internal memory. This memory is also known as the state of the network. Figure \ref{fig:RNNunrolled} gives an illustration of how RNNs could be considered as feedforward networks unrolled over time.											

\begin{figure*}[!t]
\centering
\includegraphics[width=6in]{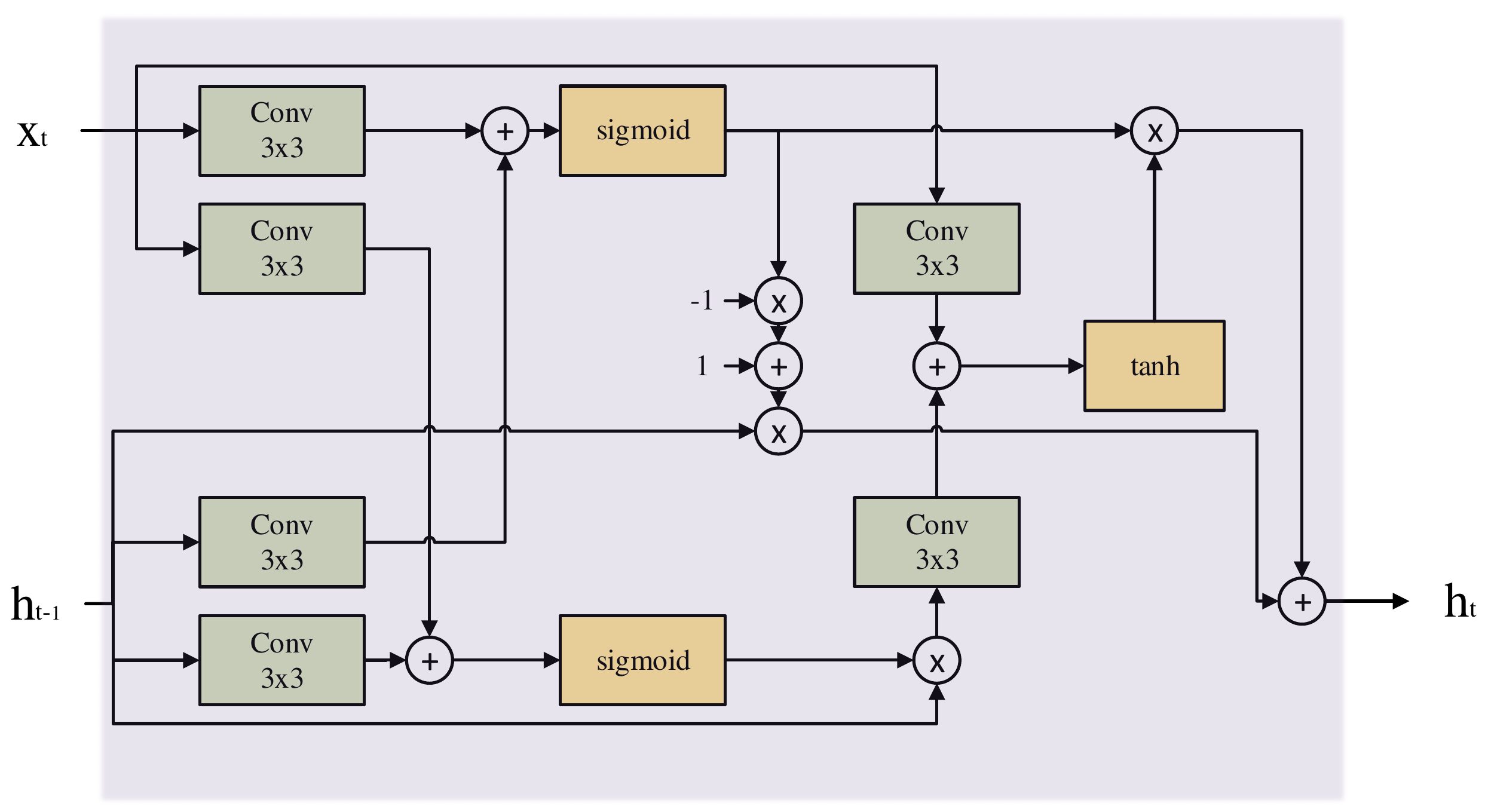}
\caption{Convolutional GRU block.}
\label{fig:ConvGRU}
\end{figure*}
\subsection{Recurrent Neural Networks}
In general, the network in each time step accepts the input of that specific slot as well as the output and the network state of the previous step. Based on these sets of data, the current state is updated and a new output is generated. In practice, the network needs to remember the state throughout the time span. In order to pass the state to future time slots, the most convenient way is to project and update the network state in each step using the following equation:
\begin{equation}
h_t=\phi(Wx_t+Uh_{t-1})
\label{EQ:RNN1}
\end{equation}
wherein $h_t$ and $x_t$ are the network state and input at time step $t$, $h_{t-1}$ is the state at the time step $t-1$, $W$ and $U$ are input and state parameters and $\phi$ is the activation function which is usually $sigmoid$ or $tanh$.\\

The biggest issue with this approach is what is known as vanishing gradients. The reason behind the vanishing gradient is the fact that when the network gets deeper in time, $h_t$ in the equation \ref{EQ:RNN1} vanishes due to successive multiplication of the $sigmoid$ or $tanh$ activation function. This causes the network to lose the information of remote states. This problem has been solved by introducing the Long Short Term Memories (LSTM) \cite{LSTM} in the mid-90s which takes advantage of gated memories in designing the block. The gated memory helps the network to forget the irrelevant information in each state and pass the important states from previous time slots without causing the vanishing gradient issue.  The LSTM blocks consist of a cell state $c$, a hidden state $h$ which is also known as the block output, and three gate operations: input gate, output gate and forget gate.\\
 
Forget gate decides on how much information from the previous cell state should be passed/ignored while updating the current cell state. The input gate is responsible for how much of the input signal contributes to the current cell state.  Finally, the hidden state (output) of the LSTM cell is updated based on the current state of the network and the output gate.
The LSTM cell is given by:
\begin{align}
&f_t = \sigma_g(W_fx_t+U_fh_{t-1}+b_f)\\
&i_t = \sigma_g(W_ix_t+U_ih_{t-1}+b_i)\\
&o_t = \sigma_g(W_ox_t+U_oh_{t-1}+b_o)\\
&c_t = f_t \odot c_{t-1}+i_t\odot \sigma_c(W_cx_t+U_ch_{t-1}+b_c)\\
&h_t = o_t \odot\sigma_c(c_t)
\end{align}
wherein $f_t$, $i_t$, and $o_t$ are the forget gate, input gate and output gate values, $c_t$ and $h_t$ are the unit's internal state and output at time step $t$, respectively. $W$, and $U$ are the learnable parameters of each equation and $b$ is is the bias. $\sigma_g$ and $\sigma_c$ are $sigmoid$ and $tanh$ functions respectively. $\odot$ is Hadamard multiplication, and $x_t$ is the input at time step $t$.\\

Gated Recurrent Units (GRU) \cite{GRU} are another implementation of RNNs which has been presented in 2014 and is a simpler implementation of LSTM. It consists of an input gate and a forget gate. The cell state and hidden state are merged as a single variable. This RNN block contains fewer parameters compared to LSTM while providing similar results \cite{GRU2}. GRU is similar to LSTM in the sense that it consist of forget gate but not an output gate. In the current work, the GRU implementation is investigated in CT reconstruction use case. The GRU cell is given by:
\begin{align}
&z_t = \sigma_g(W_zx_t+U_zh_{t-1}+b_z)\\
&r_t = \sigma_g(W_rx_t+U_rh_{t-1}+b_r)\\
&k_t = U_h(r_t\odot h_{t-1}+b_h)\\
&h_t = (1-z_t)\odot h_{t-1}+z_t \odot \sigma_h(W_hx_t)+k_t
\end{align}
wherein $z_t$ and $r_t$ are known as update and reset gate vectors. $x_t$ is the input at time step $t$ and $h_t$ is the unit's output. $W$, and $U$ are the learnable parameters of each equation and $b$ is the bias. $\sigma_g$ and $\sigma_h$ are $sigmoid$ and $tanh$ functions respectively, and $\odot$ is Hadamard multiplication.

\begin{figure*}[!t]
\centering
\includegraphics[width=5.5in]{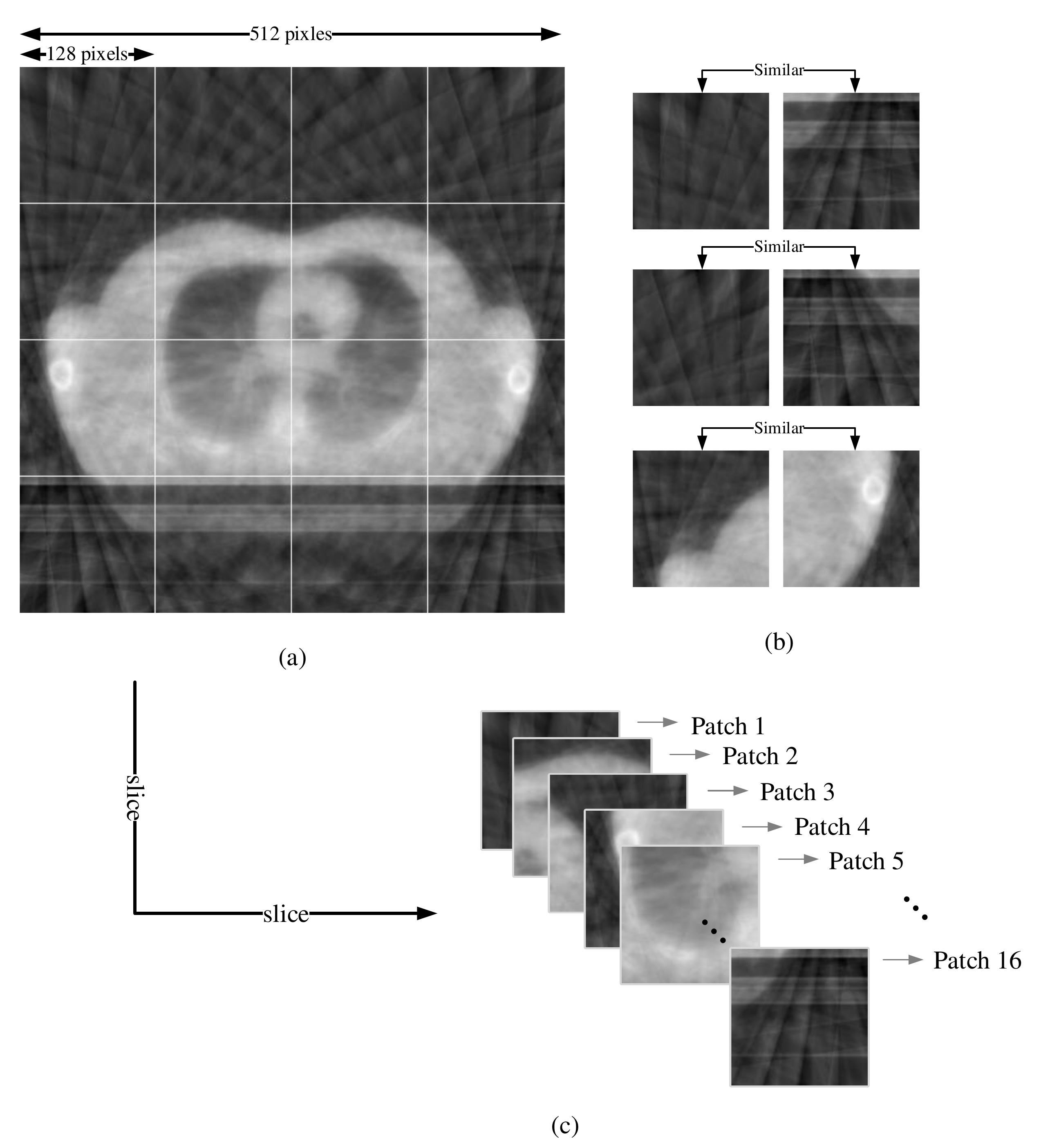}
\caption{a) The input image is divided into smaller parts. b) The streaking artefact is similar in different blocks. c) The patches are placed as a sequence to be given to an RNN. }
\label{fig:similarities}
\end{figure*}

\subsection{RNNs for Image Processing}
The conventional GRU and LSTM units are designed to process text data and not images. Giving images to these units without applying any changes causes two main issues:
since these units exploit fully connected layers in their structure, in the image processing use cases the network will contain a large number of parameters which is challenging to train and susceptible to overfitting. The other issue is that turning the images into 1-dimensional vectors loses their spatial and structural information. In order to overcome these problems, the authors in \cite{CONVRNN} proposed a convolutional GRU wherein the dot products in GRU equations are replaced with convolutional operations. This unit is represented by:
\begin{align}
&z_t = \sigma_g(W_z \ast x_t+U_z \ast h_{t-1}+b_z)\\
&r_t = \sigma_g(W_r\ast x_t+U_r\ast h_{t-1}+b_r)\\
&k_t = U_h\ast(r_t\odot h_{t-1}+b_h)\\
&h_t = (1-z_t)\odot h_{t-1}+z_t \odot \sigma_h(W_h\ast x_t)+k_t
\end{align}
wherein $\ast$ is a 4D convolutional operation mapping a 3D tensor to another 3D tensor. Note that in the current implementation all the convolutions are using $3\times3$ kernels. 

Figure \ref{fig:ConvGRU} illustrates a convolutional GRU which accepts the input at time step $t$ and the state/output of the previous unit and returns the state/output at time step $t$.

\subsection{Prepare data for RNN}
\label{sec:DataRNN}
The modern recurrent units are able to choose to remember any representative information in their current time step and pass it to the output or the next time slot. This gives the opportunity of incorporating remote time slot information in any further steps and also forget any noncooperative signal throughout the process. \\

\begin{figure}[!t]
\centering
\includegraphics[width=3.5in]{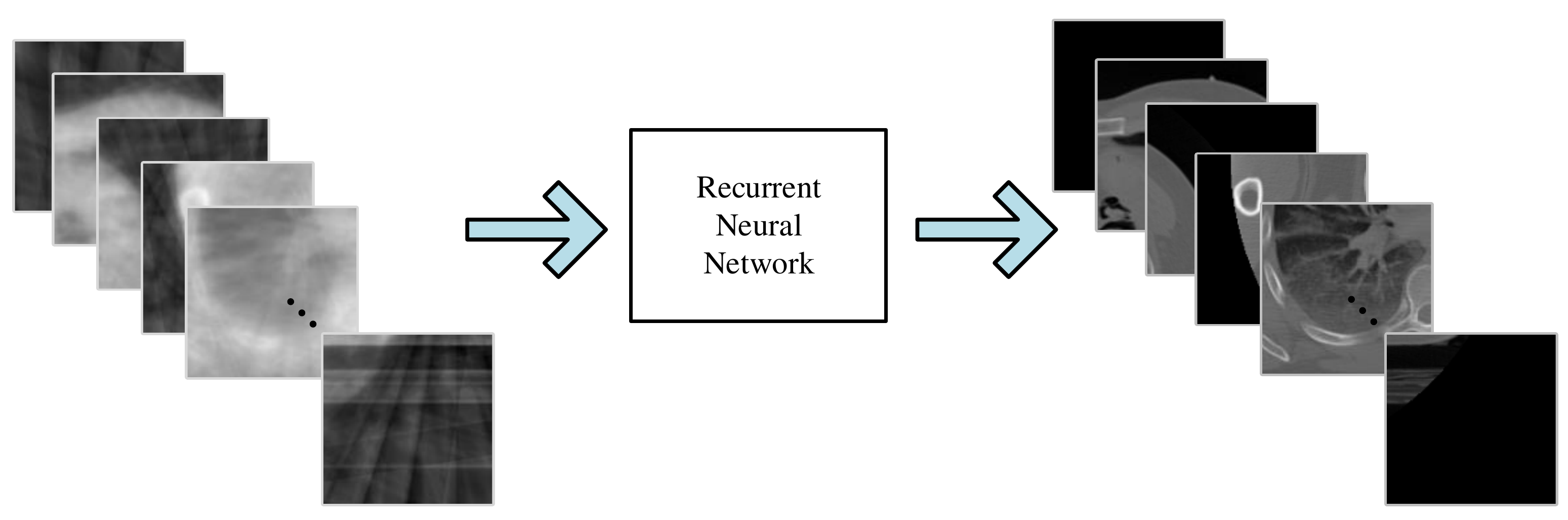}
\caption{RNN scheme to restore low dose CT image to a high quality CT image.}
\label{fig:RNNscheme}
\end{figure}

The observations on the limited angle CT reconstruction scenarios show the similarities between the streaking artefacts in different parts of the image. Figure \ref{fig:similarities} illustrates this matter wherein there are obvious similarities between the streaks in different patches. This gives the justification of using the recurrent units to learn and remove these undesired features. Figure \ref{fig:similarities} shows the proposed technique to prepare the data stream to be compatible with the RNN's pipeline. In other words, the image is sliced into several patches and each of these patches is considered as the network input in a specific time slot. For example, if the image size is $512\times 512$, and it is sliced into $128\times 128$ patches, then the designed RNN needs to accept $16$ time slots and returns $16$ outputs (Figure \ref{fig:RNNscheme}). The idea behind this design is that the network learns to remove the artifacts from early patches and transfers the knowledge to further patches since they are affected by a similar type of artefact.

\section{Network Design}
\label{sec:NetDesign}
Deep neural networks contain different types of processing units also known as layers. These layers are convolution, deconvolution, pooling, unpooling, or fully connected units. There are also regularization actions throughout the network such as drop-out \cite{dropout}, and batch normalization \cite{BATCHNORM} which act as regularization terms in the training stage and also skipped connections which facilitate the information flow inside the network. The fully convolutional deep neural networks are a subset of DNNs wherein no fully connected layers exist in the network architecture. These networks are widely used in problems where the input and the output of the network are both images, such as segmentation, depth estimation, and denoising applications. In the current work, the presented network is a fully convolutional recurrent neural network which is inspired by a fully convolutional feed-forward network known as MultiScale Dense (MSD) neural network presented in \cite{MSDorig} explained in the following section.

\begin{figure}[!h]
\centering
\includegraphics[width=3.5in]{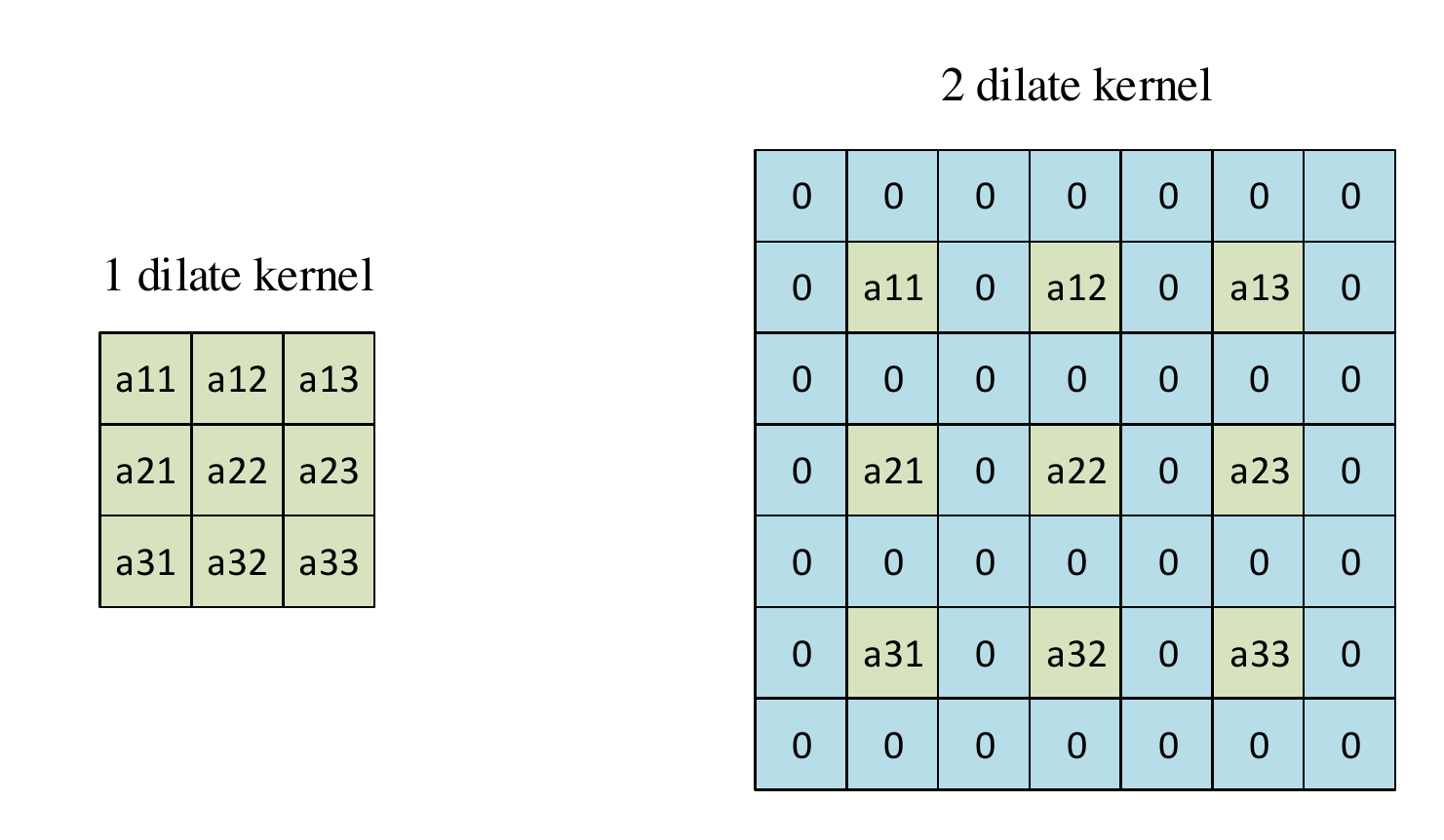}
\caption{$3 \times 3$ kernels. Left: 1 dilate. Right: 2 dilate.}
\label{fig:Dilate}
\end{figure}

\begin{figure*}[!t]
\centering
\subfloat[MSD Deep Neural Network architecture for the feed forward network]{\includegraphics[width=7in]{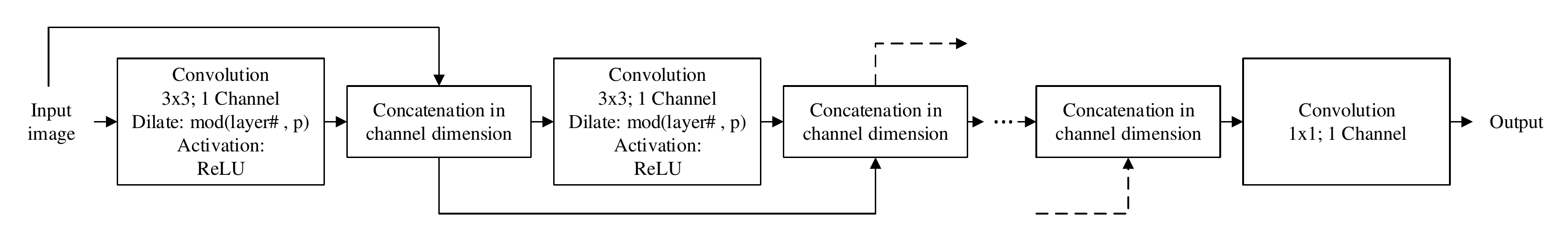}%
\label{fig:MSD}}
\vfil
\subfloat[RNN counterpart of the MSD architecture. Each arrow represents a patch of the input image]{\includegraphics[width=7in]{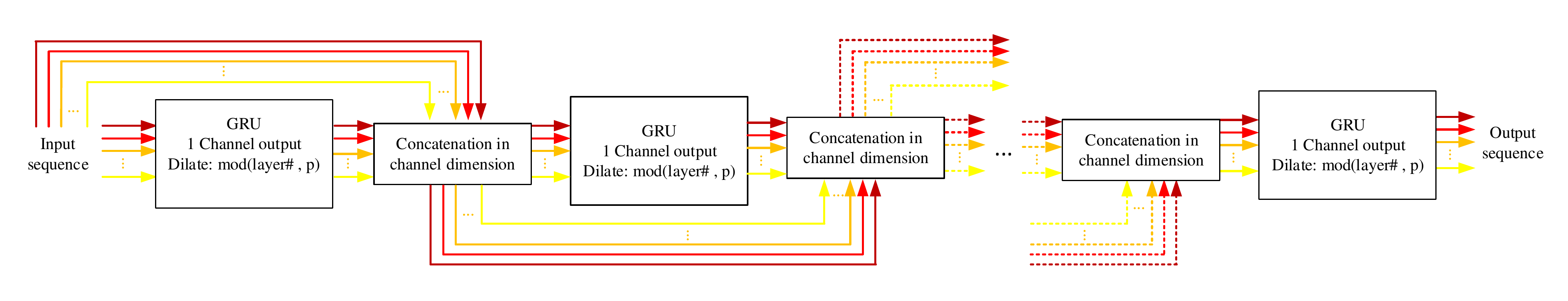}%
\label{fig:GRUMSD}}
\caption{Feedforward and RNN networks designed to remove streaking artefacts from low dose CT images.}
\label{fig:NetworksArchitecture}
\end{figure*}

\subsection{MultiScale Dense DNN}

The MSD network is a fully convolutional DNN wherein each layer accepts input from all previous layers and returns a single channel output. This network takes advantage of the dilated convolution operation. In this type of convolution, the kernel is expanded by the insertion of zero in between its parameters. For example figure \ref{fig:Dilate} shows a 1-dilated and 2-dilated $3\times 3$ convolutional kernel. The main advantage of this technique is to increase the field of view of the kernel without increasing the number of the parameters which causes overfitting and/or adding pooling layers which induce blurring to the final results.\\

The MSD network is shown in figure \ref{fig:MSD} wherein all the layers use $3\times 3$ convolutional kernels and the dilation of each kernel is specified by the layer number. The ReLU \cite{RELU} activation function is applied after each convolution. The last layer is a $1\times 1$ convolutional layer with no nonlinearity \cite{pelt2018improving}. This network proved to be effective in removing the streaking artefacts from the limited angle CT reconstructions. In \cite{pelt2018improving}, this network has been applied to FBP reconstructed images and in \cite{MYTIP1} it has been used alongside the SIRT method in an iterative manner. In the current work, this network is used to remove streaking artefact from SIRT reconstruction. The results are used to compare to the RNN approach.

\subsection{MultiScale Dense RNN}

Inspired by the feed-forward MSD, in the current work an RNN MSD has been designed wherein each layer consists of a GRU block instead of the convolutional units used in the original MSD.  This architecture is shown in figure \ref{fig:GRUMSD}. In this model, each of the GRU units takes advantage of the dilated convolution as well. In other words, the convolutional layers inside the GRU (see figure \ref{fig:ConvGRU}) exploit the same dilation operation as presented in feed-forward MSD. Given a $512\times 512$ image as the original input, based on the discussion in section \ref{sec:DataRNN}, each GRU processes 16 images of size $128\times 128$ and returns the same number of images at its output. These images are then concatenated with the output of every previous layer in the channel dimension and fed to the next GRU. It is worthwhile to mention that in the practical implementation of the RNN, the input data and also the signal between layers are 5-dimensional tensors in the order (sample\#, patch, channel, height, width).

\section{Database}
\label{sec:Database}
In the machine learning and deep learning communities, it is common to divide a single database into three subsets: Training, Validation and Test. The first two are used in the training stage and the last one is used for inference in testing. Since all the subsets are taken from a single parent database and share the same property distributions, the generalization of the trained network is not completely investigated \cite{MYFACIAL} in the testing stage. In the current work, the inter database testing method is utilized in order to provide a realistic overview of the evaluations. In this approach, networks are trained and validated on a database known as a training database and the testing is performed on a different database knows as a test database. Databases used in our work are explained below:

\begin{itemize}
\item {\bf Train Database}: The database used in the training and validation stages is the National Cancer Institutes Clinical Proteomic Tumor Analysis Consortium Pancreatic Ductal Adenocarcinoma (CPTAC-PDA) which consists of more than 45000 Pancreas images from CPTAC phase 3 patients. Images are different sizes which are all resized to $512 times 512$ for the current study. Several imaging modalities exist in this database including CT and MRI from 45 radiology and 77 pathology subjects.
\item {\bf Test Database}: For testing purposes, the Visible Human Project CT Datasets was used. This dataset contains more than 2900 CT images from 10 cases. It consists of samples from the ankle, head, hip, knee, pelvis, and shoulder from both male and female subjects. Images are $512\times 512$  therefore no resizing was applied.
\end{itemize}

In order to simulate the limited angle reconstructions, the following geometry was considered:

\begin{itemize}
\item {\bf Beam}: Parallel beam.
\item {\bf Projections}: 20 equiangular projections between 0 and 180 degrees.
\item {\bf Number of detectors}: 512.
\item {\bf Projector type}: Linear.
\item {\bf Reconstruction method}: SIRT, 100 iterations.
\end{itemize}

The geometry definition, forward projection, and SIRT reconstruction were conducted using ASTRA Toolbox \footnote{https://www.astra-toolbox.com/} \cite{astra1,astra2,astra3} for both train and test sets.

\section{Training}
\label{sec:training}
\begin{figure*}[!h]
\centering
\includegraphics[width=7in]{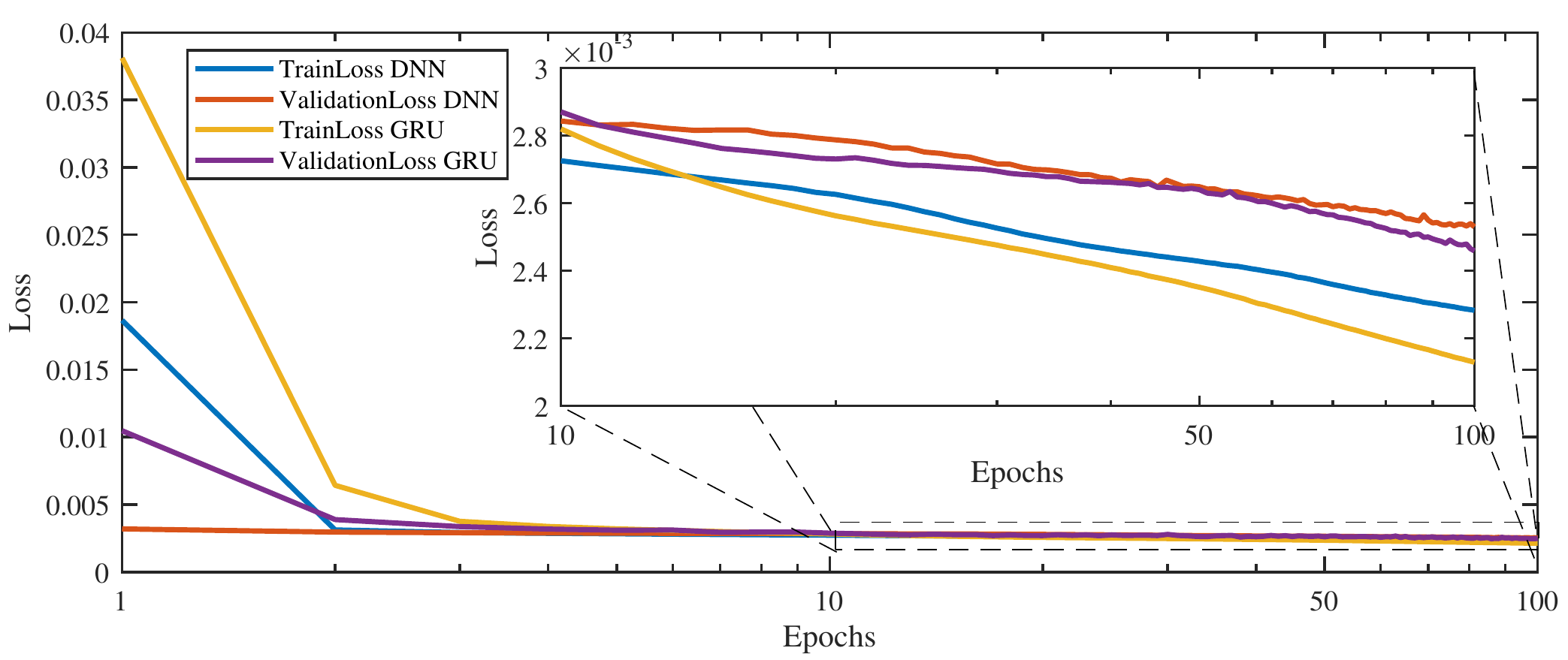}
\caption{Train and validation losses for GRU based and feedforward DNN.}
\label{fig:TrainValidationLoss}
\end{figure*}

The designed feed-forward MSD in this work exploits 15 convolutional layers with dilate range p=5. In other words, the dilation value is increased linearly in each layer and reset to 1 in every 5 layers. The RNN is designed in the same way wherein 15 GRUs are used in the design. And the same dilate range p=5 is also used in this model. Both networks are initialized uniformly in the range [-0.25,0.25], and the Adam optimized \cite{ADAM} with learning rate, $\beta_1$, $\beta_2$ and $\epsilon$ equal to $0.0001$, $0.9$, $0.999$, and $10^{-8}$, respectively was used to update the network parameters. Both networks were trained using a single GPU of an Nvidia DGX station. MXNET 1.3.0 \footnote{https://mxnet.apache.org/} \cite{MXNET}, was used on python 2.7 as the deep learning framework for training. The batch size 5 was used in the training and both models were trained for 150 epochs. The Mean Square Error given below is used as the loss function, returning a distance between the output of the network with its corresponding ground truth.

\begin{equation}
Loss = \frac{1}{B_s H W}\sum_{k=1}^{B_s}\sum_{j=1}^{H}\sum_{i=1}^{W}\big(O(i,j,k)-t(i,j,k)\big)^2\quad,
\label{eq:MSE}
\end{equation}
where $W$, $H$, and $B_s$ are the width, height, and the batch size of the input signal, respectively. The training set was divided into two subsets wherein $80\%$ of the samples were used for training and $20\%$ for validation. The training and validation losses for both feed-forward and RNN are shown in figure \ref{fig:TrainValidationLoss}. The original MSD network is called DNN and the RNN is referred to as GRU. At the early stages of training, the feedforward network shows faster convergence compared to the RNN, but in the later epochs, the RNN gives smaller loss value for the training set. The validation losses of both networks are very close to each other while RNN gives marginally better validation loss.  

\section{Results}
\label{sec:results}
The common approach in testing a learning-based method widely accepted and used in the literature is to divide the database into train, validation and test subsets. Since the network is trained and tested on the same database, the generalization of the learning-based methods is not completely investigated.   In order to provide an honest comparison between model-based and learning-based methods and also between learning-based methods, the inter-database evaluation method has been employed. In this approach, the neural networks are trained on a specific database and tested on a different dataset. This method accounts for any biasing on particular data distribution.  The train and test datasets are discussed in section \ref{sec:Database}. \\

\begin{figure}[!h]
\centering
\includegraphics[width=2.7in]{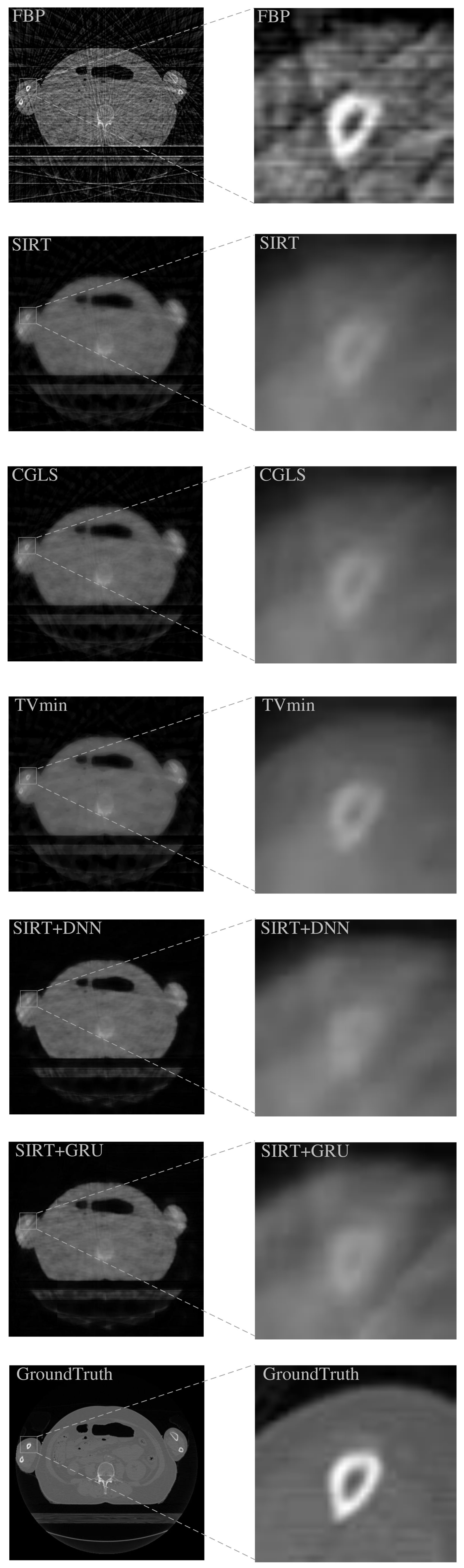}
\caption{Reconstructions from different methods. Last row illustrates the corresponding groundtruth.}
\label{fig:Results1}
\end{figure}

\begin{figure}[!h]
\centering
\includegraphics[width=2.7in]{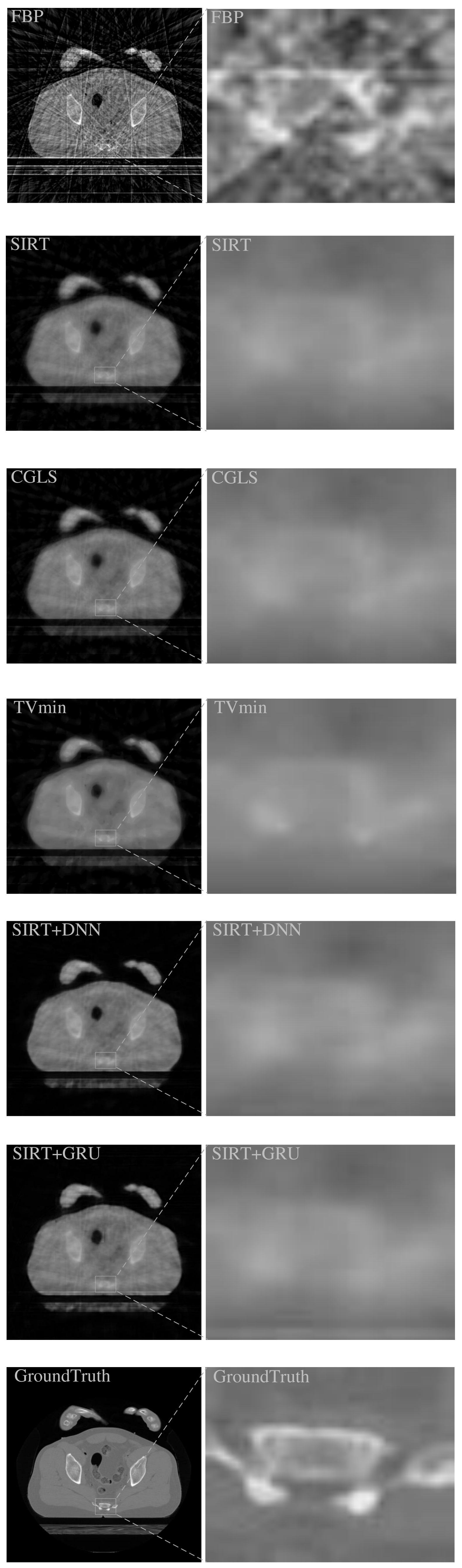}
\caption{Reconstructions from different methods. Last row illustrates the corresponding ground truth.}
\label{fig:Results2}
\end{figure}

In this section, the output of the trained feed-forward and RNN networks are evaluated on 10 different subsets of the test dataset and the visual and numerical results are provided alongside with model-based methods FBP, SIRT, CGLS, and TVmin. The sinogram noise evaluations are also presented in section \ref{sec:NoiseEval}.

\subsection{Numerical and Visual Results}
\label{sec:NumAndVisResults}
In this section, the results of the noiseless simulations are compared between different model-based and learning-based methods on the test set. The numerical results are provided using Peak Signal to Noise Ratio (PSNR), Mean Squared Error (MSE) and Structural Similarity Index (SSIM) \cite{SSIM} measurements. Tables 1 and 3 correspond to the PSNR and MSE measurements.  The TVmin and CGLS methods return the best values for these two measurements amongst the model-based methods. Since PSNR and MSE return pixel-level correctness and ignore the structural information, we used the SSIM to evaluate the structural correctness between different reconstruction methods. In table 2 the SSIM measurements are shown for different methods. In this table, the TVmin method provides marginally better structural correctness compared to CGLS and iterative methods return higher scores compared to FBP. \\

Considering learning-based algorithms the feedforward network gives marginally better measurements compared to the recurrent approach for PSNR, SSIM, and MSE. In general, the learning-based methods provide higher quality numerical measurements compared to the model-based methods.\\
    
Figures \ref{fig:Results1} and \ref{fig:Results2} provide visual results for all methods. The FBP reconstruction returns the sharpest results while it highly suffers from the streaking artefact. The SIRT and CGLS methods produce less streaking artefact but at the same time, they introduce blurring to the reconstruction. The result of the TVmin method provides the sharpest output but most of the details are entangled into a cartoon shape artefact. The SIRT+DNN method which is the feedforward network applied to the SIRT output is able to remove a large amount of streaking artefact but suffers from the loss of fine details due to blurring. The SIRT+GRU approach successfully removes the streaking artefact from SIRT and the results are sharper than SIRT+DNN method. 
\begin{table*}
\centering
\ra{1.4}
\begin{tabular}{@{}rcccccccccc@{}}\toprule
& Ankle F & Head F & Hip F & Knee F & Pelvis F & Shoulder F & Head M & Hip M & Pelvis M & Shoulder M\\
\toprule
{\bf Model based}&&&&&&&&&&\\
{FBP}&27.7$\pm$0.7&27.7$\pm$1.8&25.7$\pm$1.3&27.7$\pm$0.5&23.2$\pm$0.2&24.3$\pm$1.0&26.2$\pm$1.2&24.8$\pm$0.7&23.4$\pm$0.7&23.4$\pm$1.2\\
{SIRT}&36.7$\pm$0.7&35.4$\pm$1.8&34.6$\pm$0.9&36.6$\pm$0.6&32.6$\pm$0.4&33.8$\pm$0.8&35.1$\pm$1.6&34.0$\pm$0.5&32.3$\pm$0.5&32.5$\pm$1.1\\
{CGLS}&{37.0$\pm$0.6}&35.6$\pm$1.8&34.9$\pm$0.8&36.8$\pm$0.6&{32.9$\pm$0.4}&{34.1$\pm$0.9}&{35.4$\pm$1.6}&34.3$\pm$0.4&32.6$\pm$0.5&32.7$\pm$1.0\\
{TVmin}&36.9$\pm$0.7&{35.7$\pm$1.8}&{35.0$\pm$0.9}&{36.9$\pm$0.5}&32.8$\pm$0.4&34.0$\pm$0.8&35.4$\pm$1.7&34.4$\pm$0.5&{32.7$\pm$0.5}&{32.8$\pm$1.1}\\
{\bf Learning based}&&&&&&&&&&\\
{SIRT+DNN}&{\bf 39.2$\pm$0.5}&{\bf 37.1$\pm$1.8}&{\bf 36.2$\pm$1.2}&{\bf 38.7$\pm$0.7}&{\bf 34.1$\pm$0.5}&{\bf 35.1$\pm$0.8}&{\bf 37.0$\pm$1.6}&{\bf 35.8$\pm$0.4}&{\bf 34.5$\pm$0.4}&{\bf 34.6$\pm$1.0}\\
{SIRT+GRU}&38.9$\pm$0.4&37.0$\pm$1.7&35.9$\pm$1.0&38.3$\pm$0.6&33.9$\pm$0.4&34.9$\pm$0.8&36.8$\pm$1.5&35.5$\pm$0.4&34.2$\pm$0.3&34.4$\pm$0.9\\
\toprule
\end{tabular}
\caption{PSNR}
\end{table*}

\begin{table*}
\centering
\ra{1.4}
\begin{tabular}{@{}rcccccccccc@{}}\toprule
& Ankle F & Head F & Hip F & Knee F & Pelvis F & Shoulder F & Head M & Hip M & Pelvis M & Shoulder M\\
\toprule
{\bf Model based}&&&&&&&&&&\\
FBP&0.33$\pm$3e-2&0.27$\pm$7e-2&0.29$\pm$1e-2&0.30$\pm$2e-2&0.26$\pm$2e-2&0.29$\pm$4e-2&0.27$\pm$6e-2&0.29$\pm$1e-2&0.27$\pm$2e-2&0.27$\pm$3e-2\\
SIRT&0.76$\pm$2e-2&0.72$\pm$6e-2&0.73$\pm$2e-2&0.76$\pm$2e-2&0.68$\pm$1e-2&0.72$\pm$2e-2&0.73$\pm$5e-2&0.70$\pm$1e-2&0.64$\pm$2e-2&0.65$\pm$3e-2\\
CGLS&{0.78$\pm$3e-2}&{0.74$\pm$6e-2}&0.74$\pm$2e-2&0.77$\pm$2e-2&{0.70$\pm$1e-2}&0.73$\pm$3e-2&0.73$\pm$5e-2&0.72$\pm$2e-2&0.66$\pm$2e-2&0.67$\pm$3e-2\\
TVmin&{0.78$\pm$2e-2}&{0.74$\pm$6e-2}&{0.75$\pm$2e-2}&{0.78$\pm$2e-2}&{0.70$\pm$1e-2}&{0.74$\pm$3e-2}&{0.74$\pm$5e-2}&{0.74$\pm$2e-2}&{0.67$\pm$2e-2}&{0.68$\pm$3e-2}\\
{\bf Learning based}&&&&&&&&&&\\
SIRT+DNN&{\bf 0.91$\pm$7e-3}&{\bf 0.92$\pm$9e-3}&{\bf 0.89$\pm$1e-2}&{\bf 0.91$\pm$4e-3}&{\bf 0.85$\pm$7e-3}&{\bf 0.86$\pm$1e-2}&{\bf 0.90$\pm$1e-2}&{\bf 0.89$\pm$6e-3}&{\bf 0.86$\pm$6e-3}&{\bf 0.86$\pm$1e-2}\\
SIRT+GRU&0.89$\pm$8e-3&0.91$\pm$1e-2&0.87$\pm$2e-2&0.90$\pm$5e-3&0.83$\pm$9e-3&0.85$\pm$1e-2&0.88$\pm$2e-2&0.88$\pm$6e-3&0.85$\pm$7e-3&0.85$\pm$2e-2\\
\toprule
\end{tabular}
\caption{SSIM}
\end{table*}

\begin{table*}
\centering
\ra{1.4}
\begin{tabular}{@{}rcccccccccc@{}}\toprule
& Ankle F & Head F & Hip F & Knee F & Pelvis F & Shoulder F & Head M & Hip M & Pelvis M & Shoulder M\\
\toprule
{\bf Model based}&&&&&&&&&&\\
FBP&6e-3$\pm$1e-3&7e-3$\pm$3e-3&1e-2$\pm$4e-3&6e-3$\pm$7e-4&2e-2$\pm$1e-3&1e-2$\pm$3e-3&9e-3$\pm$2e-3&1e-2$\pm$2e-3&2e-2$\pm$2e-3&2e-3$\pm$4e-3\\
SIRT&8e-4$\pm$1e-4&{1e-3$\pm$4e-4}&{1e-3$\pm$3e-4}&{8e-4$\pm$1e-4}&{2e-3$\pm$2e-4}&{1e-3$\pm$3e-4}&{1e-3$\pm$4e-4}&{1e-3$\pm$2e-4}&{2e-3$\pm$3e-4}&{2e-3$\pm$5e-4}\\
CGLS&{7e-4$\pm$1e-4}&{1e-3$\pm$4e-4}&{1e-3$\pm$2e-4}&{8e-4$\pm$1e-4}&{2e-3$\pm$2e-4}&{1e-3$\pm$3e-4}&{1e-3$\pm$4e-4}&{1e-3$\pm$2e-4}&{2e-3$\pm$3e-4}&{2e-3$\pm$4e-4}\\
TVmin&8e-4$\pm$1e-4&{1e-3$\pm$4e-4}&{1e-3$\pm$2e-4}&{8e-4$\pm$1e-4}&{2e-3$\pm$2e-4}&{1e-3$\pm$3e-4}&{1e-3$\pm$4e-4}&{1e-3$\pm$2e-4}&{2e-3$\pm$2e-4}&{2e-3$\pm$5e-4}\\
{\bf Learning based}&&&&&&&&&&\\
SIRT+DNN&{\bf 4e-4$\pm$6e-5}&{\bf 8e-4$\pm$3e-4}&{\bf 9e-4$\pm$2e-4}&{\bf 5e-4$\pm$8e-5}&{\bf 1e-3$\pm$1e-4}&{\bf 1e-3$\pm$2e-4}&{\bf 8e-4$\pm$3e-4}&{\bf 1e-3$\pm$1e-4}&{1e-3$\pm$1e-4}&{\bf 1e-3$\pm$3e-4}\\
SIRT+GRU&5e-4$\pm$5e-5&{\bf 8e-4$\pm$3e-4}&1e-3$\pm$2e-4&{\bf 5e-4$\pm$8e-5}&{\bf 1e-3$\pm$1e-4}&{\bf 1e-3$\pm$2e-4}&{\bf 8e-4$\pm$2e-4}&{\bf 1e-4$\pm$1e-4}&{\bf 1e-3$\pm$1e-4}&{\bf 1e-3$\pm$3e-4}\\
\toprule
\end{tabular}
\caption{MSE}
\end{table*}


\subsection{Evaluations with acquisition noise}
\label{sec:NoiseEval}

\begin{figure}[!h]
\centering
\includegraphics[width=3.1in]{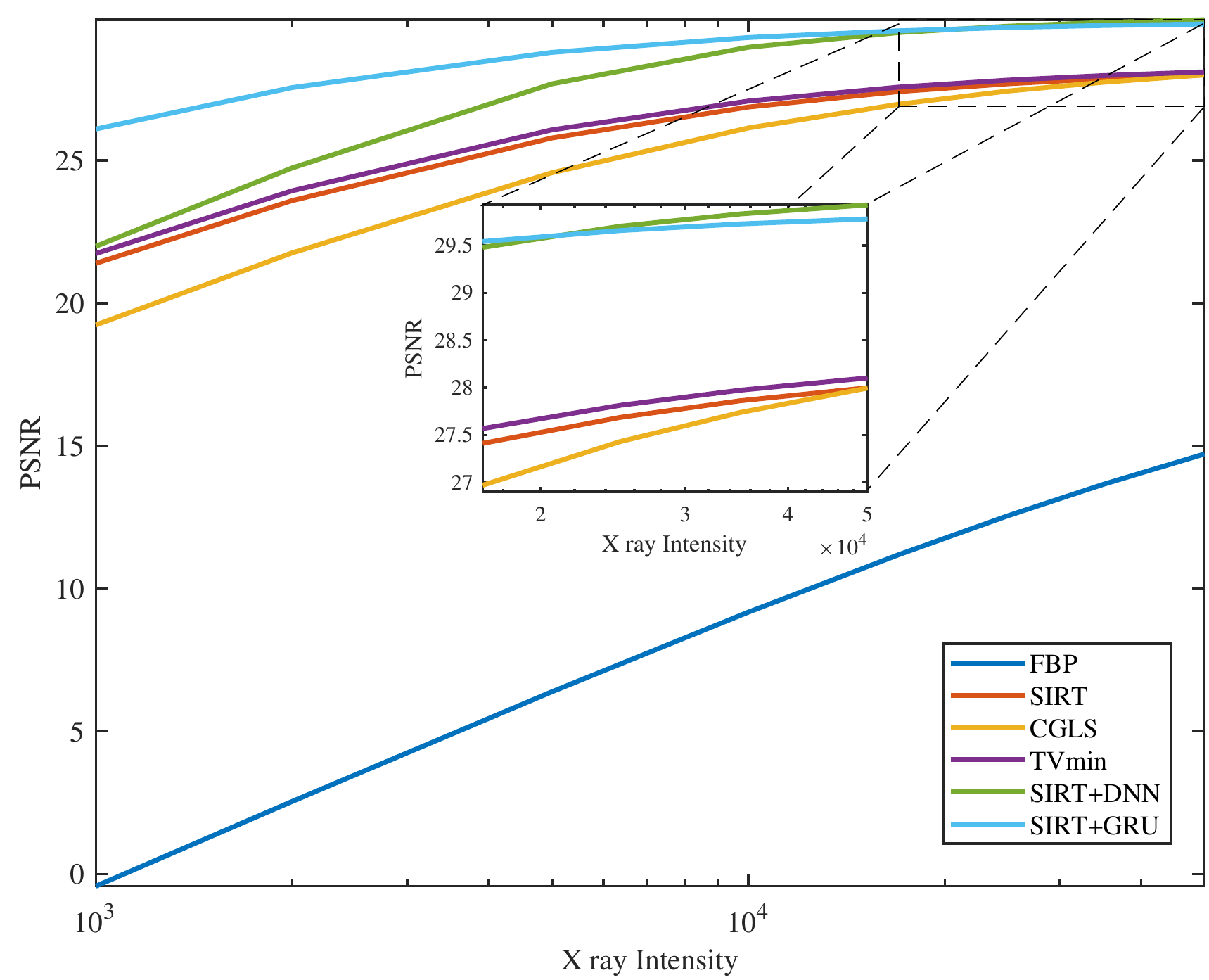}
\caption{Mean PSNR wrt X ray intensity for different methods.}
\label{fig:NoisePSNR}
\end{figure}

\begin{figure}[!h]
\centering
\includegraphics[width=3.1in]{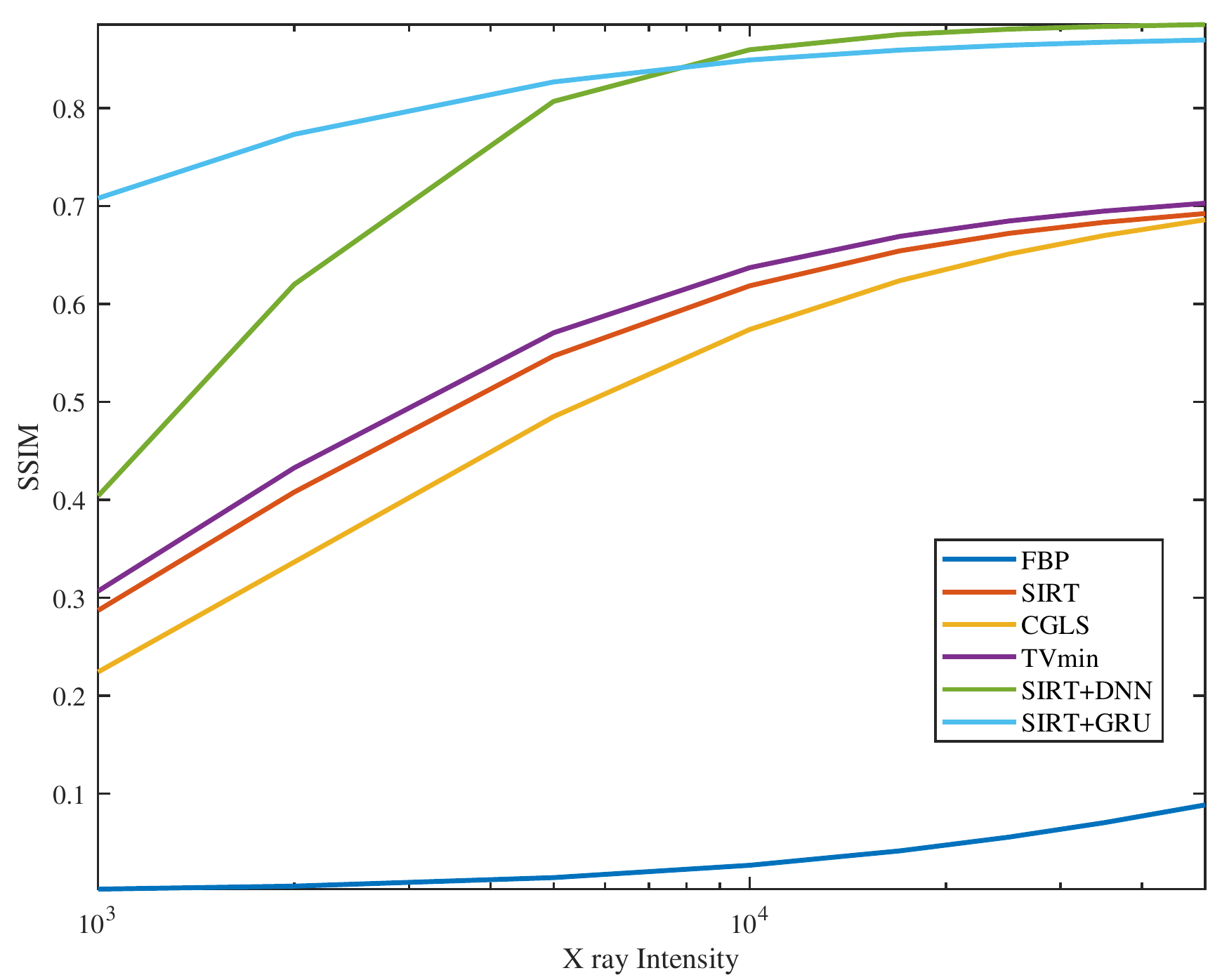}
\caption{Mean SSIM wrt X ray intensity for different methods.}
\label{fig:NoiseSSIM}
\end{figure}

\begin{figure}[!h]
\centering
\includegraphics[width=3.1in]{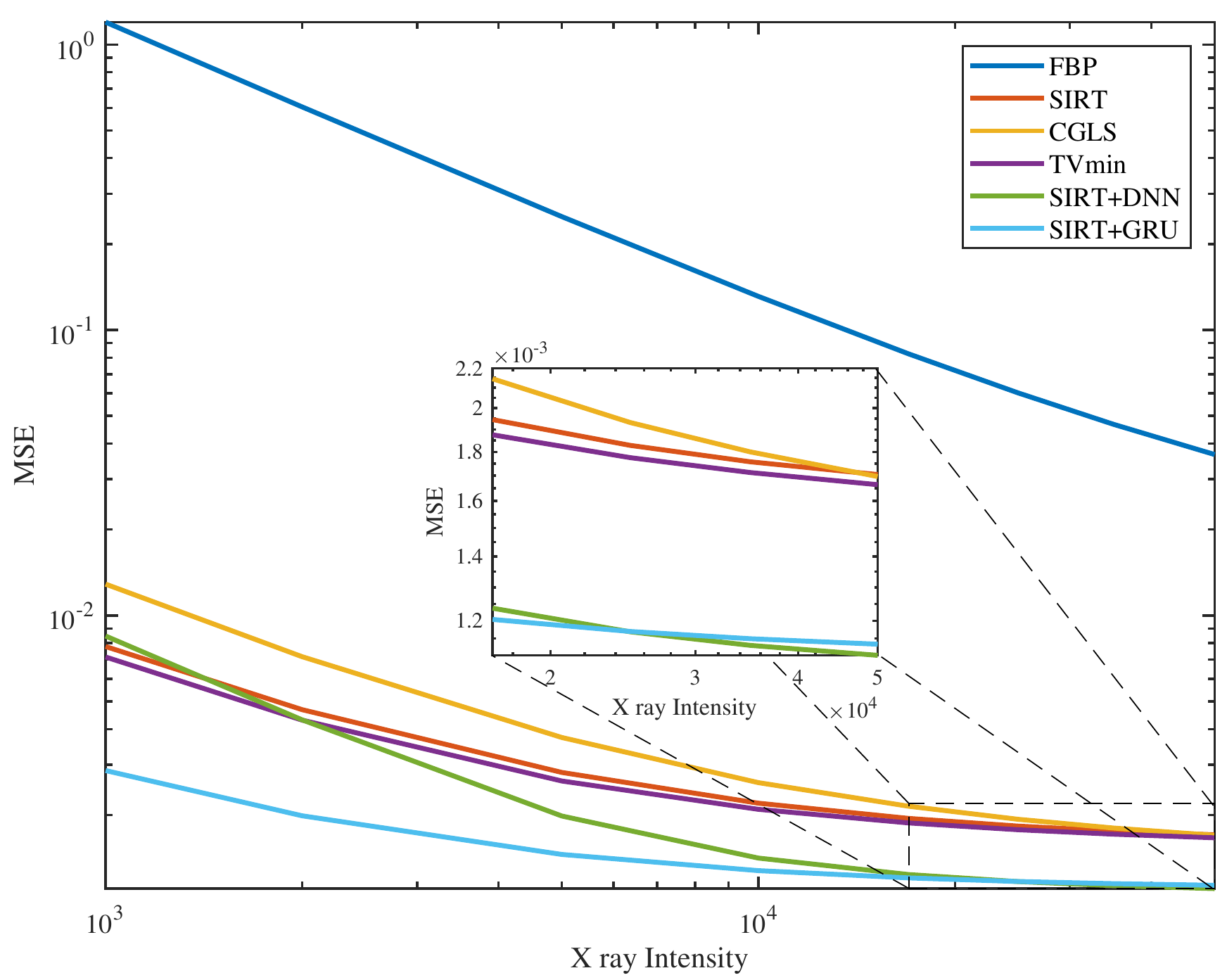}
\caption{Mean MSE wrt X ray intensity for different methods..}
\label{fig:NoiseMSE}
\end{figure}

In the current work, the noise realization is accomplished by adding a Poisson distribution to the sinogram signal. This is implemented in the ASTRA toolbox wherein the noise is added using the photon count information. In other words, the noise is realized based on the number of photons interacting with each detector pixel. This is also known as X-ray intensity.  In this work, this intensity is varied between 1000 and 50000 photons per pixel. The lower the X-ray intensity the higher the noise and vice versa. The performance of different reconstruction methods with noise are shown in figures \ref{fig:NoisePSNR} to \ref{fig:NoiseMSE}. In these figures, the horizontal axis represents the X-ray intensity and the vertical axis shows the PSNR, SSIM and MSE measurements. The values are the average between all test samples in all test sets.\\

It has been shown that in the presence of sinogram noise, the learning-based methods return higher quality numerical results compared to the model-based approaches. Between learning-based methods, in high X-ray intensities, the feedforward network gives better results compared to the recurrent model. But in high noise cases, the recurrent method manages to return more honest reconstructions especially for the SSIM measurement which is related to the structural consistency compared to the ground truth.

\subsection{Considerations on processing power}
Each GRU block consists of six convolutional layers, three of which accept the input from that specific time slot and other three accept the signal with the shape of the previous block's output. In the current implementation, the number of the input channels increases with the depth of the network as explained in section \ref{sec:NetDesign}, and the output of each GRU block is a single-channel image.\\

Consider the image of the size $S_1 \times S_2$ with $c$ channels which is divided into $b_1 \times b_2$ blocks. Each GRU block consists of 
\begin{equation}
N_{P_{GRU}} = 3^3 \times c + 3^3
\end{equation}
parameters compared to the feedforward network convolutional layer which consists of  
\begin{equation}
N_{P_{DNN}} = 3^2\times c
\end{equation}
parameters.
At the same time, each GRU unit at each timeslot has 

\begin{equation}
N_{M_{GRU}} = \frac{S_1}{b_1}\times\frac{S_2}{b_2}\times 3^3 \times c + \frac{S_1}{b_1}\times\frac{S_2}{b_2}\times 3^3
\end{equation}

multiplications in the convolution operations, compared to

\begin{equation}
N_{M_{GRU}} = S_1 \times S_2 \times 3^2 \times c
\end{equation}

for the feedforward convolutional layer. At each time step, the GRU block needs less processing power compared to the feedforward network. But after unrolling the recurrent unit, there are $b_1 \times b_2$ GRU units which result in 

\begin{equation}
N_{M_{GRUunrolled}} = S_1\times S_2 \times 3^3 \times c + S_1 \times S_2 \times 3^3
\end{equation}

multiplications in the convolution operations for the recurrent implementation. In other words, the number of multiplication with respect to the number of input channels grows ${(b_1\times b_2)}/{3}$ times faster for feed-forward network compared to the single time slot GRU and grows 3 times faster for the unrolled GRU compared to the feedforward network.
Considering the similar outcomes of both methods (see section \ref{sec:NumAndVisResults}) in low noise cases, the GRU method is able to be implemented as $b_1 \times b_2$ separate blocks, each contains ${(b_1\times b_2)}/{3}$ times fewer multiplications than the feed-forward method. In the use cases where the unrolled recurrent network is calculated as a single bock, the feed-forward network needs almost three times fewer resources. 

\section{Conclusions}
\label{sec:conslusions}
In this article, the use of RNN in image restoration for low dose CT use cases has been investigated and results have been compared to corresponding feedforward network and model-based reconstruction algorithms. The approach is to divide the input image into several patches and use each patch as a timeslot input of the RNN. The network learns the streaking at early patches and applies it to the next parts of the input image. The inter-database testing approach has been applied to the network wherein the train and test databases are gathered from different sources. The results show similar numerical outcomes for RNN compared to the feedforward network and slightly sharper reconstructions for RNN. The processing power is also investigated which shows that the RNN contains almost 3 times more multiplications compared to the feedforward network, while there are more opportunities for serialization in RNN in the case of low processing power hardware. Considering the noisy input, the RNN gives better numerical results in high noise conditions where most of the methods fail to provide reasonable outputs. 
\section*{Acknowledgment}
This work is financially supported by VLAIO (Flemish Agency for Innovation and Entrepreneurship), through the ANNTOM project HBC.2017.0595.\\

\ifCLASSOPTIONcaptionsoff
  \newpage
\fi

\bibliographystyle{IEEEtran}
\bibliography{bibmag,bibrnn,ShababCH1,ShababCH3,bib}

\end{document}